\documentclass[12pt]{article}
\usepackage{amsmath}
\usepackage[english]{babel}
\usepackage{graphicx}

\begin{document}

\title{Hill's nano-thermodynamics is equivalent with Gibbs' thermodynamics
for curved surfaces}
\author{Dick Bedeaux\thanks{%
corresponding author email: dick.bedeaux@chem.ntnu.no} and Signe Kjelstrup \\
Porelab, Department of Chemistry, \\
Norwegian University of Science and Technology, NTNU, Trondheim, Norway}
\maketitle

\begin{abstract}
We review first how properties of curved surfaces can be studied using
Hill's thermodynamics, also called nano-thermodynamics. We proceed to show
for the first time that Hill's analysis is equivalent to Gibbs for curved
surfaces. This simplifies the study of surfaces that are curved on the
nano-scale, and opens up a possibility to study non-equilibrium systems in a
systematic manner.
\end{abstract}

\section{Introduction}

To master transport on the nano-scale, say in catalysis, for electrode
reactions or for fluid transport in porous media, is of great importance
[1,2,3]. But systems on this scale do not have additive thermodynamic
properties [4,5,6], and this makes their thermodynamic description
complicated. Hill [4,5] devised a scheme to obtain thermodynamics properties
at equilibrium for this scale in the early 1960'ies. He showed that the
thermodynamics on this scale was crucially modified, a fact that may have
hampered the further development since then. Nevertheless, he provided a
systematic basis, which is also the first necessary step in a development of
a non-equilibrium thermodynamics theory. We believe that Hill's method is
better suited, to make progress in the direction of non-equilibrium. To
facilitate its use, it may then be useful to make it better anchored in the
more familiar thermodynamic description of Gibbs. This communication
addresses how this can be done for curved surfaces, a central topic in
nano-scale physics.

Gibbs [7] gave a thermodynamic theory of equilibrium surfaces; a theory that
was extended by Tolman [8] and Helfrich [9]. Tolman [8] introduced what is
now called the Tolman-length, while Helfrich [9] gave an expansion to second
order in the curvature and introduced two bending rigidities and the natural
curvature. Blokhuis and Bedeaux [10-12] derived statistical mechanical
expressions for these coefficients.

Hill chose a different route to this problem with his small system
thermodynamics. He introduced an ensemble of small systems, and used the
replica energy to obtain thermodynamic properties that depended on the
surface area and curvatures.

Both Hill and Gibbs described how to account for thermodynamic contributions
from curved surfaces. This implies that Hill's analysis should reproduce the
description given by Gibbs. Hill [4, page 168] only addressed this issue for
the special case of a spherical drop in a super saturated vapor. We have
recently verified that this is generally true for a flat surface [13]. In
this letter we verify this property for curved surfaces. Doing this, we
verify that the properties of curved surfaces on the nanoscale can be
studied also by Hill's method.

This has at least two consequences. First, we can better understand why a
relatively new method, the small system method [13,14], can be used to
produce properties of the surface, even if there is no direct study of
surfaces in the method. We are able to conclude that all information, even
of the surface of a system, can be obtained from the fluctuations of the
number of particles and the energy in the small system. Second, it supports
Hill's idea to deal with nano-scale systems as ensembles. This we expect
will facilitate a derivation of non-equilibrium properties like the entropy
production and flux-force relations [15].

Before we study the equivalence of Hill's method with Gibbs', we give a
short repetition of the essential points of Hill's method. We then discuss
the curvature dependencies in detail. Doing this, we hope to revitalize
Hill's work [4,5] and contribute to the contemporary need for a more
systematic nano-scale thermodynamics away from equilibrium.

\section{The idea of Hill's method}

\begin{figure}[tbp]
\centering
\includegraphics[width= 0.85\textwidth]{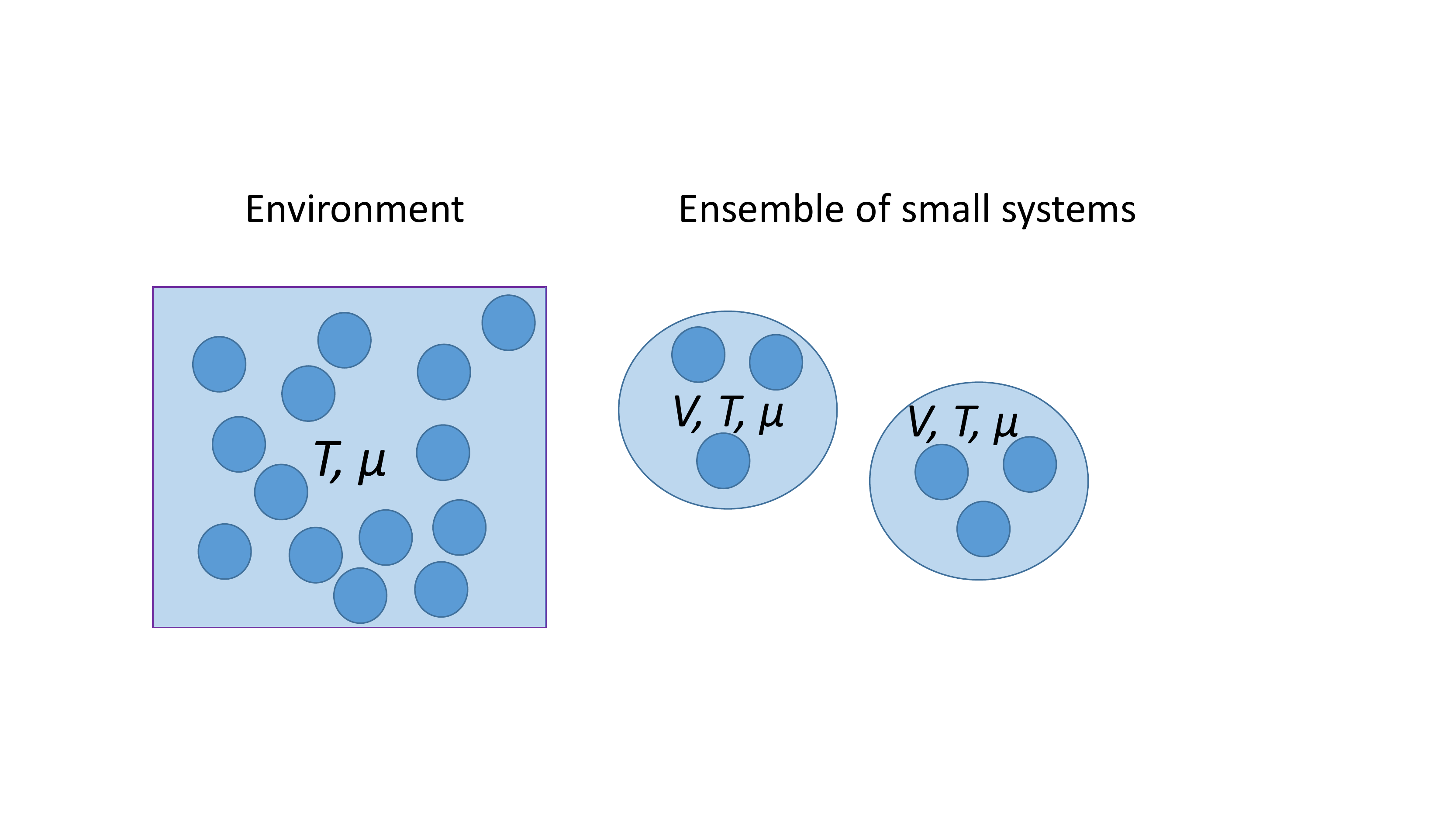}
\caption{An example of an ensemble of small systems (only two replicas with
a volume $V$ are shown) in contact with the environment. The ensemble
exchanges energy and mass with the heat- and particle-baths of the
environment without being in direct contact.}
\label{fig:CG_Small_System}
\end{figure}

We recapitulate the essentials of Hill's method. Consider a small system
with volume $V$\ in contact with the environment of temperature $T$ and
chemical potential $\mu $. The system can exchange heat and particles with
the environment. An ensemble of replicas is now constructed by considering $%
\mathcal{N}$ independent, distinguishable small systems, characterized by $T$%
, $V$, $\mu $. Figure \ref{fig:CG_Small_System} shows two replicas in
contact with the heat- and particle-bath of the environment. The environment
defines $T$ and $\mu $.

The idea is now that the \textit{ensemble} of small systems follows the laws
of macroscopic thermodynamic systems when $\mathcal{N}$ is large enough. The
Gibbs equation for the ensemble is given by 
\begin{equation}
dU_{t}=TdS_{t}-p\mathcal{N}dV+\mu dN_{t}+Xd\mathcal{N}  \label{1}
\end{equation}%
where $U_{t}$, $S_{t}$, $N_{t}$ are the total internal energy, entropy and
number of particles of the whole grand-canonical ensemble, which are
functions of the environmental variables ($T$,$V$,$\mu $) and $\mathcal{N}$.
The subscript $t$ denotes properties of the full ensemble. Furthermore $p$
is the pressure.\ The so-called replica energy of an ensemble member is now
given by 
\begin{equation}
X(T,V,\mu )\equiv \left( \frac{\partial U_{t}}{\partial \mathcal{N}}\right)
_{S_{t},V,N_{t}}  \label{2}
\end{equation}%
The replica energy, $X\equiv -\hat{p}V$, can be interpreted as the work
required to increase the volume of the ensemble by adding one ensemble
member, while $p\mathcal{N}dV$ is the work required to increase the volume
of the ensemble by increasing the volume of each member. The $T,\ p$ and $%
\mu $\ satisfy equations similar to Eq.\ref{2}. Note that $V_{t}=\mathcal{N}%
V $.

Unlike in the thermodynamic limit, the thermodynamics of small systems
depends on the choice of the environmental control variables [4,5,6]. For
other ensembles, like e.g. the canonical $(T,V,N)$-ensemble, the
thermodynamic properties differ from the ones in Fig.1. In the thermodynamic
limit one can use Legendre transformations to go from one choice to another.
This is not possible for small systems. We restrict ourselves to a
one-component fluid. The extension to a mixture is straightforward.

Using Euler's theorem of homogeneous functions of degree one, we integrate
Eq.\ref{1}, holding $T$,$V$,$\mu $ and $X$ constant, and obtain 
\begin{equation}
U_{t}(T,V,\mu ,\mathcal{N})=TS_{t}(T,V,\mu ,\mathcal{N})-\hat{p}(T,V,\mu )V%
\mathcal{N}+\mu N_{t}(T,V,\mu ,\mathcal{N})  \label{3}
\end{equation}%
where we have used the definition $X\equiv -\hat{p}V$. The ensemble averages
of internal energy, particle number, and entropy are given by%
\begin{eqnarray}
U_{t}(T,V,\mu ,\mathcal{N}) &\equiv &\mathcal{N}U(T,V,\mu )  \notag \\
N_{t}(T,V,\mu ,\mathcal{N}) &\equiv &\mathcal{N}N(T,V,\mu )  \notag \\
S_{t}(T,V,\mu ,\mathcal{N}) &\equiv &\mathcal{N}S(T,V,\mu )  \label{4}
\end{eqnarray}%
While $U\ $and $N\ $fluctuate because the small systems are open, the
entropy does not, and is the same for each ensemble member [4]. By
substituting the relations in Eq.\ref{4} into Eq.\ref{3}, we can write the
average energy of a single small system as 
\begin{equation}
U(T,V,\mu )=TS(T,V,\mu )-\hat{p}(T,V,\mu )V+\mu N(T,V,\mu )  \label{5}
\end{equation}%
For ease of notation we omit from now the dependence on $T,V,\mu $ and $%
\mathcal{N}$. In the thermodynamic limit $p=\hat{p}$, and we are left with
the classical thermodynamic relation. The last term is necessary in a
one-component system because the chemical potential depends on $V, N$.

We obtain the Gibbs relation for the small system by introducing the
relations in Eq.\ref{4} into Eq.\ref{1}, using $X\equiv -\hat{p}V$ and Eq.%
\ref{5} 
\begin{equation}
dU=TdS-pdV+\mu dN  \label{6}
\end{equation}%
By differentiating Eq.\ref{5} and combining the result with Eq.\ref{6}, the
Gibbs-Duhem-like equation becomes 
\begin{equation}
d(\hat{p}V)=SdT+pdV+Nd\mu  \label{7}
\end{equation}%
from which we can derive the following expression for a small system 
\begin{equation}
p=\left( \frac{\partial \hat{p}V}{\partial V}\right) _{T,\mu }=\hat{p}%
+V\left( \frac{\partial \hat{p}}{\partial V}\right) _{T,\mu }  \label{8}
\end{equation}%
This identity promoted Hill [4] to give the variable $\hat{p}$ the name 
\emph{integral} pressure. The variable $p$ was then called the \emph{%
differential} pressure. As long as the systems are so small that the
integral pressure $\hat{p}$ depends on the volume, the differential pressure 
$p$ differ from the integral pressure. Expressions similar to Eq.\ref{8}
apply for $S$\ and $N$.

This illustrates the framework developed by Hill [4]; the framework that
allows us to systematically handle the thermodynamics of small systems, see
the original work for more details.

\section{Curvature dependency by Hill's method}

A property that is extensive in the thermodynamic limit can always be
written as the sum of this limit plus a small size correction 
\begin{equation}
A(T,V,\mu )\equiv a(T,V,\mu )V\equiv a^{\infty }(T,\mu )V+a^{s}(T,\mu
,c_{1},c_{2})\Omega  \label{9}
\end{equation}%
where $a(T,V,\mu )$ is the density of $A(T,V,\mu )$ per unit of volume. In
the second equality we used that thermodynamic limit value $a^{\infty
}(T,\mu )\equiv A^{\infty }(T,V,\mu )/V$ is independent of the volume. We
shall build on the fact that the thermodynamic limit value is uniquely
determined, meaning that Eq.\ref{9} determines $a^{s}$\ uniquely.

The density $a$\ and the whole $A$\ do not only depend on the size $V$\ of
the volume, but also on its shape. This dependence was not explicitly
indicated in Eqs.\ref{1} to \ref{8}. The shape-dependence is the reason why
there is a small size-correction $a^{s}(T,\mu ,c_{1},c_{2})$, which depends
on the principle curvatures, $c_{1}\equiv 1/R_{1}$ and $c_{2}\equiv 1/R_{2}$
where $R_{1}$\ and $R_{2}$ are the principle radii of curvature of the
surface of the small volume. The symbol $\Omega $ is the surface area. The
principal curvature of a surface is illustrated for simplicity for a
two-dimensional case in Fig.2. The small sphere, touching the wall of the
small (ellipsoidal) system, defines the radius of curvature $R_1$.

The principle curvatures will generally vary along the surface of a small
volume, cf. Fig.2. They are only constant when the system is a sphere or a
cylinder. This implies that one should everywhere on the surface use the
local values and integrate the corresponding contributions over the surface,
see Helfrich [9]. We will here take $c_{1}$ and $c_{2}$ constant, which
simplifies the analysis considerably. A generalization to a varying $c_{1}$
and $c_{2}$ can be done, and does not alter the result.

\begin{figure}[tbp]
\centering
\includegraphics[width= 0.85\textwidth]{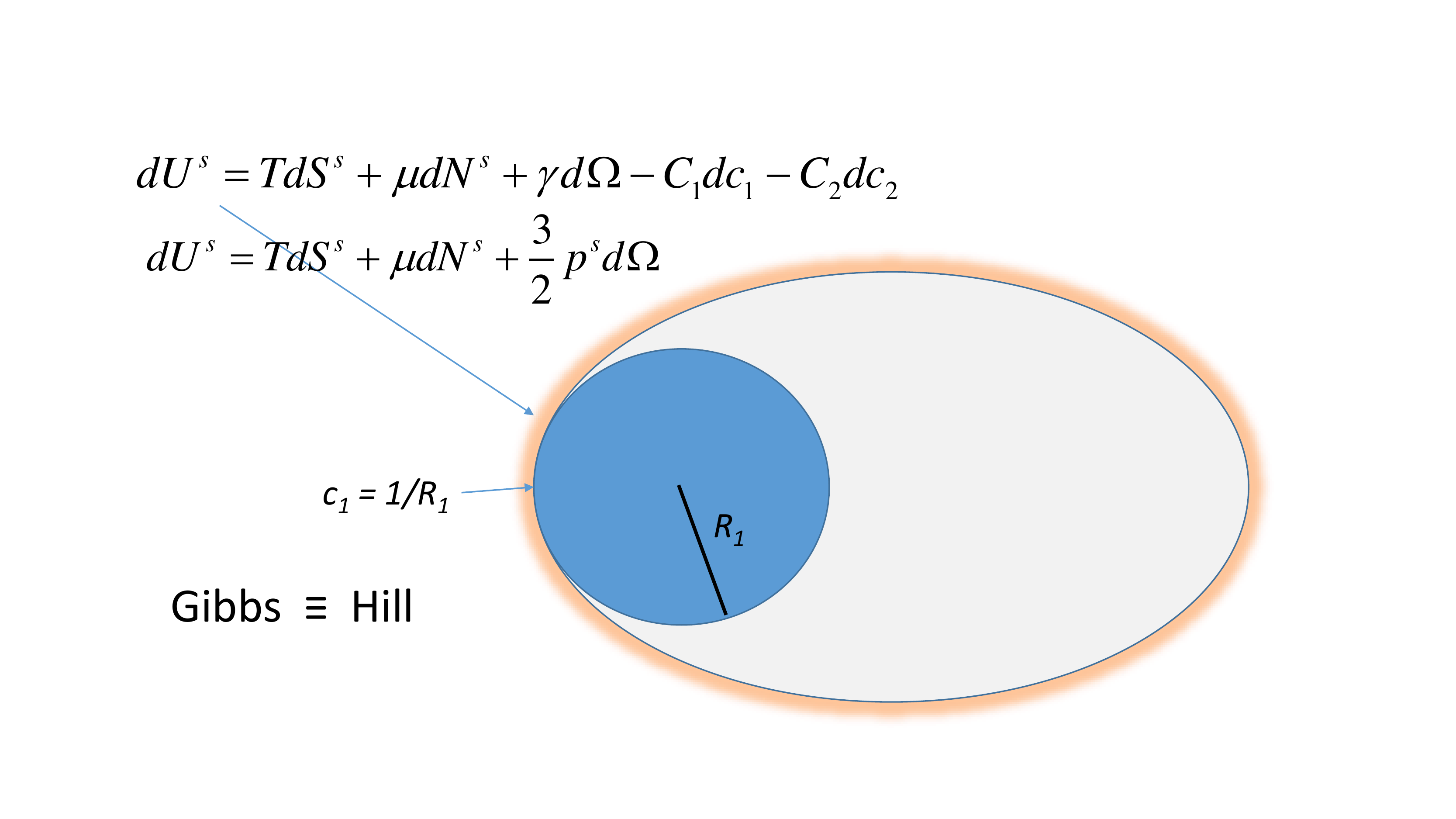}
\caption{The radius of curvature of a two-dimensional surface (disk). Gibbs
thermodynamics for the surface energy is equivalent to Hill's for a small
system.}
\label{fig:Curvature2}
\end{figure}

Eq.\ref{9} is valid for $U,S$ and $N$. For the pressures of the volume we
have 
\begin{eqnarray}
p(T,V,\mu )V &=&p^{\infty }(T,\mu )V+p^{s}(T,\mu ,c_{1},c_{2})\Omega  \notag
\\
\hat{p}(T,V,\mu )V &=&\hat{p}^{\infty }(T,\mu )V+\hat{p}^{s}(T,\mu
,c_{1},c_{2})\Omega  \label{10}
\end{eqnarray}%
It follows from Eq.\ref{10} together with Eq.\ref{8} that%
\begin{equation}
p^{\infty }(T,\mu )=\hat{p}^{\infty }(T,\mu )  \label{11}
\end{equation}%
In the thermodynamic limit the small-size corrections are negligible. The
Euler equation, Eq.\ref{5}, becomes using Eq.\ref{11} 
\begin{equation}
u^{\infty }=Ts^{\infty }-p^{\infty }+\mu n^{\infty }  \label{12}
\end{equation}%
The Gibbs equation, Eq.\ref{6}, becomes%
\begin{equation}
dU^{\infty }=TdS^{\infty }-p^{\infty }dV+\mu dN^{\infty }  \label{13}
\end{equation}%
By using also Eq.\ref{11}, Gibbs-Duhem Eq.\ref{7} becomes 
\begin{equation}
dp^{\infty }=s^{\infty }dT+n^{\infty }d\mu  \label{14}
\end{equation}%
Not surprisingly these relations have their usual form. This is because they
apply to the thermodynamic limit. Subtracting Eq.\ref{12} times $V$\ from Eq.%
\ref{5} and dividing the result by $\Omega $\, we obtain the Euler relation
for small-size corrections%
\begin{equation}
u^{s}(T,\mu ,c_{1},c_{2})=Ts^{s}(T,\mu ,c_{1},c_{2})-\hat{p}^{s}(T,\mu
,c_{1},c_{2})+\mu n^{s}(T,\mu ,c_{1},c_{2})  \label{15}
\end{equation}%
By subtracting Eq.\ref{13} from Eq.\ref{6} we obtain for small-size
corrections%
\begin{equation}
dU^{s}(T,\mu ,c_{1},c_{2})=TdS^{s}(T,\mu ,c_{1},c_{2})-p^{s}(T,\mu
,c_{1},c_{2})\frac{\Omega }{V}dV+\mu dN^{s}(T,\mu ,c_{1},c_{2})  \label{16}
\end{equation}%
From the definitions $L\equiv V^{1/3}$ and $\Omega /V\equiv c_{s}/L$ we have%
\begin{eqnarray}
\left( \frac{\partial \Omega }{\partial V}\right) _{T,\mu } &=&\left( \frac{%
\partial \Omega }{\partial L}\right) _{T,\mu }\left( \frac{\partial L}{%
\partial V}\right) _{T,\mu }=\left( 2c_{s}L\right) \left( \frac{1}{3L^{2}}%
\right) =\frac{2}{3}\frac{c_{s}}{L}=\frac{2}{3}\frac{\Omega }{V}  \notag \\
&\Longrightarrow &\frac{\Omega }{V}dV=\frac{3}{2}d\Omega  \label{17}
\end{eqnarray}%
We used as condition that the change of the volume did not imply a change in
shape. By substituting the last expression into Eq.\ref{16}, we obtain
Gibbs' equation that applies when small-size contributions are relevant 
\begin{equation}
dU^{s}=TdS^{s}-\frac{3}{2}p^{s}d\Omega +\mu dN^{s}  \label{18}
\end{equation}%
The Gibbs-Duhem equation for systems with small system corrections similarly
becomes 
\begin{equation}
d\left( \hat{p}^{s}(T,\mu ,c_{1},c_{2})\Omega \right) =S^{s}(T,\mu
,c_{1},c_{2})dT+\frac{3}{2}p^{s}(T,\mu ,c_{1},c_{2})d\Omega +N^{s}(T,\mu
,c_{1},c_{2})d\mu  \label{19}
\end{equation}
In order to compare with Gibbs results (below), we use Eq.\ref{8} which gives%
\begin{equation}
p\left( T,V,\mu \right) =\frac{\partial \hat{p}\left( T,V,\mu \right) V}{%
\partial V}=\frac{\partial \left( \hat{p}^{\infty }\left( T,\mu \right)
V\right) }{\partial V}+\frac{\partial \left( \hat{p}^{s}(T,\mu
,c_{1},c_{2})\Omega \right) }{\partial V}  \label{20}
\end{equation}%
This results in%
\begin{equation}
p^{s}\left( T,\mu ,c_{1},c_{2}\right) \frac{\Omega }{V}=\hat{p}^{s}(T,\mu
,c_{1},c_{2})\frac{\partial \Omega }{\partial V}+\Omega \frac{\partial \hat{p%
}^{s}(T,\mu ,c_{1},c_{2})}{\partial c_{1}}\frac{\partial c_{1}}{\partial V}%
+\Omega \frac{\partial \hat{p}^{s}(T,\mu ,c_{1},c_{2})}{\partial c_{2}}\frac{%
\partial c_{2}}{\partial V}  \label{21}
\end{equation}%
Both curvatures change when $V$\ changes. As in the derivation of Eq.\ref{17}%
\, we find%
\begin{equation}
\frac{\partial c_{1}}{\partial V}=-\frac{c_{1}}{3V}\text{ \ and \ }\frac{%
\partial c_{2}}{\partial V}=-\frac{c_{2}}{3V}  \label{22}
\end{equation}%
Again, the change of the volume did not imply a change in shape. By
introducing this equation and Eq.\ref{17}\, we obtain the small system
pressure 
\begin{equation}
p^{s}\left( T,\mu ,c_{1},c_{2}\right) =\frac{2}{3}\hat{p}^{s}(T,\mu
,c_{1},c_{2})-\frac{c_{1}}{3}\frac{\partial \hat{p}^{s}(T,\mu ,c_{1},c_{2})}{%
\partial c_{1}}-\frac{c_{2}}{3}\frac{\partial \hat{p}^{s}(T,\mu ,c_{1},c_{2})%
}{\partial c_{2}}  \label{23}
\end{equation}

\section{Comparing with Gibbs' results}

We are now in a position where we can compare the Euler equation \ref{15}
with the one given by Gibbs [7] (see Eq.502 on page 229 of his collected
works, volume 1): 
\begin{equation}
u^{s}(T,\mu ,c_{1},c_{2})=Ts^{s}(T,\mu ,c_{1},c_{2})+\gamma (T,\mu
,c_{1},c_{2})+\mu n^{s}(T,\mu ,c_{1},c_{2})  \label{24}
\end{equation}%
where $\gamma $\ is the common surface tension. It follows that small system
pressure can be identified by the surface tension: 
\begin{equation}
\gamma (T,\mu ,c_{1},c_{2})=-\hat{p}^{s}(T,\mu ,c_{1},c_{2})  \label{25}
\end{equation}%
By introducing this into Eq.\ref{23}, we obtain for the differential pressure%
\begin{equation}
p^{s}\left( T,\mu ,c_{1},c_{2}\right) =-\frac{2}{3}\gamma (T,\mu
,c_{1},c_{2})+\frac{c_{1}}{3}\frac{\partial \gamma (T,\mu ,c_{1},c_{2})}{%
\partial c_{1}}+\frac{c_{2}}{3}\frac{\partial \gamma (T,\mu ,c_{1},c_{2})}{%
\partial c_{2}}  \label{26}
\end{equation}%
By using this in the Gibbs relation, we have 
\begin{equation}
dU^{s}=TdS^{s}+\gamma d\Omega -\frac{1}{2}\frac{\partial \gamma }{\partial
c_{1}}c_{1}d\Omega -\frac{1}{2}\frac{\partial \gamma }{\partial c_{2}}%
c_{2}d\Omega +\mu dN^{s}  \label{27}
\end{equation}%
It follows from Eqs.\ref{17} and \ref{22} that%
\begin{equation}
c_{1}d\Omega =-2\Omega dc_{1}\text{ \ and \ }c_{2}d\Omega =-2\Omega dc_{2}
\label{28}
\end{equation}%
By introducing this in the Gibbs relation, we obtain 
\begin{eqnarray}
&&dU^{s}(T,\Omega ,\mu ,c_{1},c_{2})=TdS^{s}(T,\Omega ,\mu
,c_{1},c_{2})+\gamma (T,\mu ,c_{1},c_{2})d\Omega  \notag \\
&&+\frac{\partial \gamma (T,\mu ,c_{1},c_{2})}{\partial c_{1}}\Omega dc_{1}+%
\frac{\partial \gamma (T,\mu ,c_{1},c_{2})}{\partial c_{2}}\Omega dc_{2}+\mu
dN^{s}(T,\Omega ,\mu ,c_{1},c_{2})  \label{29}
\end{eqnarray}%
The corresponding Euler relation was already given in Eqs.\ref{15} and \ref%
{24}. The Gibbs-Duhem equation, \ref{19}, becomes with Eqs.\ref{26} and \ref%
{28} 
\begin{eqnarray}
d\gamma (T,\mu ,c_{1},c_{2}) &=&-s^{s}(T,\mu ,c_{1},c_{2})dT+\frac{\partial
\gamma (T,\mu ,c_{1},c_{2})}{\partial c_{1}}dc_{1}  \notag \\
&&+\frac{\partial \gamma (T,\mu ,c_{1},c_{2})}{\partial c_{2}}%
dc_{2}-n^{s}(T,\mu ,c_{1},c_{2})d\mu  \label{30}
\end{eqnarray}%
The coefficients $C_{1}$ and $C_{2}$ are now identified with 
\begin{equation}
C_{1}=\frac{\partial \gamma (T,\mu ,c_{1},c_{2})}{\partial c_{1}}\Omega 
\text{ \ and \ }C_{2}=\frac{\partial \gamma (T,\mu ,c_{1},c_{2})}{\partial
c_{2}}\Omega
\end{equation}%
These are identities which follow from Eq.\ref{29} as Maxwell relations. By
introducing these, we obtain 
\begin{eqnarray}
dU^{s}(T,\Omega ,\mu ,c_{1},c_{2})= && TdS^{s}(T,\Omega
,\mu,c_{1},c_{2})+\gamma (T,\mu ,c_{1},c_{2})d\Omega \\
&&+ C_1 dc_{1}+ C_2 dc_{2}+\mu dN^{s}(T,\Omega ,\mu ,c_{1},c_{2})  \notag
\label{29b}
\end{eqnarray}%
Equation \ref{29b} is exactly the one given by Gibbs [7] (see Eq.493 on page
225 in his collected works, volume 1) for thermodynamics of surfaces of
heterogeneous systems.

\section{Concluding remarks}

We have seen above that the analysis given by Gibbs [7] of the
thermodynamics of heterogeneous systems is equivalent to the thermodynamics
of small systems as formulated by Hill [4] 90 years later.

But Hill extended the treatment of small systems much beyond the study of
curved surfaces. He used the same ensemble procedure to study, say,
adsorption, crystallization, bubbles, all under different environmental
conditions. With the equivalence proven, we can take advantage of the
broader method of Hill in the study of curved and other surfaces. One of the
advantages of Hill's method is that one obtains the properties of the small
system, including surface and curvature contributions without the need to
immediately introduce the dividing surface.

It is interesting to note that the small system method, derived from Hill's
basis, gives information on thermodynamic properties of small systems,
without having to actually create these small systems. We have earlier
demonstrated, using this method, that a scaling law exists, relating surface
properties to properties in the thermodynamic limit [13,14]. The important
conclusion appears; that information of the surface properties is contained
in the characteristic fluctuations of for instance the number of particles
and the energy that take place in the small system. This is a very general
observation, that also supports the idea that Hill's thermodynamics may
provide a fruitful basis, also for the derivation and use of non-equilibrium
thermodynamics for the nano-scale. It is our hope that this can stimulate
similar efforts, and lead to a development of non-equilibrium
nano-thermodynamics.

\subsection*{Acknowledgment}

The authors are grateful to the Research Council of Norway through its
Centers of Excellence funding scheme, project number 262644, PoreLab.
Discussions with Edgar Blokhuis, Bj\o rn A. Str\o m and Sondre K. Schnell
are much appreciated.

\subsection*{References}

1. Richardson H. H. et al., Nano Lett. \textbf{6} (2006) 783.\newline
2. Govorov A. O. et al., Nanoscale Research Letters \textbf{1} (2006) 84.%
\newline
3. Jain P. K., El-Sayed I. H. and El-Sayed M. A., Nano Today \textbf{2}
(2007) 18.\newline
4. Hill T. L., \textit{Thermodynamics of small systems} (Dover, New York)
1994.\newline
5. Hill T. L., Perspective: Nanothermodynamics, Nano Lett. \textbf{1} (2001)
111.\newline
6. Latella I., P\'{e}rez-Madrid A., Campa A., Casetti L., Ruffo S., Phys.
Rev. Lett. \textbf{114} (2015) 230601.\newline
7. Gibbs .J W., \textit{The Scientific Papers of J. Willard Gibbs, Volume 1,
Thermodynamics} (Ox Bow Press, Woodbridge, Connecticut) 1993.\newline
8. Tolman R.C., J. Chem. Phys. \textbf{17} (1949) 333.\newline
9. Helfrich W., Z. Naturforsch. \textbf{28c} (1973) 693.\newline
10. Blokhuis E.M. and Bedeaux D., J. Chem. Phys. \textbf{95} (1991) 6986.%
\newline
11. Blokhuis E.M. and Bedeaux D., Physica A \textbf{184} (1992) 42.\newline
12. Blokhuis E.M. and Bedeaux D., Heterogeneous Chemistry Reviews, \textbf{1}
(1994) 55.\newline
13. Str\o m B.A., Simon J-M., Schnell S.K., Kjelstrup S., He J., Bedeaux D.,
PCCP \textbf{19} (2017) 9016.\newline
14. Schnell S.K., Vlugt T.J.H., Simon J-M., Bedeaux D., Kjelstrup S., Chem.
Phys. Letters \textbf{504} (2011) 199. \newline
15. Kjelstrup S. and Bedeaux D., \textit{Nonequilibrium Thermodynamics for
Heterogeneous Systems, Series on Statistical Mechanics, Vol. 16} (World
Scientific, Singapore) 2008.\newline

\end{document}